\documentstyle[epsf,amsfonts,aps]{revtex}
\begin{document}

\title{Kondo Impurity in a Mesoscopic Ring: Charge Persistent 
       Current\footnote{To appear in the Proceedings from {\em
Electron Transport in Mesoscopic Physics,} satellite to LT22 (1999),
J. Low Temp. Phys. (in press)}}

\author{H.-P.\ Eckle,$^{1,2}$ H.\ Johannesson,$^3$ and C.\ A.\ Stafford$^{2,4}$}

\address{$^1$Department of Physics, University of Jyv\"askyl\"a,
             FIN--40351 Jyv\"askyl\"a, Finland \\
         $^2$Department of Physics, University of Arizona, 
             1118 E.\ 4th Street, Tucson, AZ 85721, USA\\
         $^3$Institute of Theoretical Physics,
             Chalmers University of Technology and G\"oteborg University,
             SE 412 96 G\"oteborg, Sweden\\
	 $^4$Fakult\"at f\"ur Physik, 
             Albert-Ludwigs-Universit\"at,
             Hermann-Herder-Stra\ss e 3,
	     D-79104 Freiburg, Germany}

\maketitle

\begin{abstract}
We study the influence of a magnetic 
impurity or ultrasmall quantum dot
on the charge persistent current of a
mesoscopic ring.
The system consists of
electrons in a one-dimensional ring 
threaded by spin--dependent
Aharonov-Bohm/Casher fluxes, coupled via an antiferromagnetic
exchange interaction to a localized electron.
By passing to a basis of electron states with definite parities, the problem is 
mapped onto a Kondo model for the even-parity channel
plus free electrons in the odd-parity channel.
The twisted boundary conditions representing the fluxes
couple states of opposite parity unless the twist angles satisfy
$\phi_\alpha=f_\alpha\pi$, where $f_\alpha$ are integers, with spin index
$\alpha=\uparrow, \downarrow$. For these special values of $\phi_\alpha$,
the model is
solved exactly by a Bethe ansatz.
Special cases are investigated in detail.
In particular we show that the
charge stiffness in the case $\phi_\uparrow=\phi_\downarrow$
is insensitive to the presence of the magnetic impurity/quantum dot.

\end{abstract}

\section{INTRODUCTION}
Recent experimental breakthroughs in identifying a Kondo effect in 
a quantum dot connected to leads have led to a flurry of activities in
{\em mesoscopic Kondo physics}.\cite{review} In a parallel development,
data on Kondo scattering from a {\em single} magnetic impurity has also been
reported.\cite{li} With Kondo physics now available in the laboratory
at the mesoscopic scale, the full richness of this paradigm of many-body
physics can be explored.

One of the basic fingerprints of coherent electron transport is that of the
Aharonov-Bohm (AB) effect, which
has already been observed in microstructured conducting rings coupled to a
quantum dot.\cite{yacoby}
A natural question to ask is how the AB effect and
its dual, the Aharanov-Casher (AC) effect, in
a ring coupled to a quantum dot and hence e.g.\ the persistent currents
would be modified by the many-body 
correlations present in the Kondo regime.

As is well-known, charge and spin persistent currents are the equilibrium
responses of a multiply-connected system to a magnetic AB flux and/or
an AC flux of a charged string (ABC fluxes) piercing the 
system.\cite{by,BIL,Cheung,mbcs}
The persistent current of 
a ring coupled via tunneling to a quantum dot was investigated via
perturbation  theory and numerical diagonalization by B\"uttiker
and Stafford.\cite{mbcs} 
In this article, 
we study a variant of the problem where electrons in a one-dimensional 
(1D) ring threaded by spin-dependent ABC fluxes 
$\phi_\alpha$ ($\alpha =\uparrow, \downarrow $) are
coupled via antiferromagnetic exchange to a localized electron, representing
a magnetic impurity or quantum dot.
A detailed
analysis shows that this model can be mapped onto the integrable Kondo
model for special values of $\phi_\alpha$,
corresponding to periodic and
antiperiodic boundary conditions. 
The model is solved via Bethe ansatz for these special 
values of $\phi_\alpha$,
and it is shown that the charge stiffness for $\phi_\uparrow=\phi_\downarrow$
is insensitive to the Kondo scattering, implying that spin-charge
separation holds even on the mesoscopic scale in this model.
However, for $\phi_\uparrow\neq\phi_\downarrow$, the charge persistent current
is affected by the presence of the magnetic impurity, as can be shown by a 
detailed analysis of the Bethe ansatz equations.

\section{EXACTLY SOLVABLE MODEL}
The system we are considering
is described  in the continuum limit by the 1D
Hamiltonian,
\begin{equation}
H = - \frac{\hbar^2}{2m} \sum_{\alpha} \int_0^L dx \psi_{\alpha}^{\dagger}(x)
                                             \partial^2_x
                                             \psi_{\alpha}(x) +
    \lambda \sum_{\alpha,\beta} \psi_{\alpha}^{\dagger}(0)
                                 \vec{\sigma}_{\alpha \beta}
                                 \psi_{\beta}(0) \cdot \vec{S},
\label{eq:Hinitial}
\end{equation}
where 
$\lambda > 0$ is an antiferromagnetic Kondo coupling, $m$ is the 
electron mass, $L$ is the circumference of the ring, $\vec{S}$ is the impurity 
spin (located at $x=0$), 
and $\psi_{\alpha}$ is an electron field with spin index 
$\alpha = \uparrow, \downarrow$.
The effect of the fluxes $\phi_\alpha$ has been gauged away\cite{by}
and encoded in twisted boundary conditions:
\begin{equation}
\psi_{\alpha}(L) = e^{i \phi_\alpha}\psi_{\alpha}(0),
\label{eq:tbc}
\end{equation}
where $\phi_{\uparrow, \downarrow} = 2\pi(\Phi/\Phi_0 \pm 4\pi\tau/F_0)$.
Here $\Phi$ is the magnetic flux enclosed by the ring and $\tau$ the
line charge density of a charged string passing through the center of
the ring. $\Phi_0 = hc/e$ is the elementary magnetic flux
quantum, with $F_0=h c/\mu$ its electric analogue, $\mu$ being the projection of
the magnetic moment along the string.

We are interested in an exact solution of 
the problem described by 
(\ref{eq:Hinitial}) and (\ref{eq:tbc}), with a particular eye on 
how the Kondo interaction in (\ref{eq:Hinitial}) may affect the 
persistent current induced by the boundary phase angles $\phi_\alpha$
describing the ABC fluxes.
Since the essential physics of the system is confined to a small
region around the left and right Fermi points, we can 
linearize the quadratic dispersion in (\ref{eq:Hinitial}) around $\pm k_{\mathrm{F}}$ 
and introduce 
left ($l$) and right ($r$) moving
chiral fields:
\begin{equation}
\psi_{\alpha}(x) \sim \mbox{e}^{-ik_{\mathrm{F}} x}\psi_{l,\alpha}(x) +
                   \mbox{e}^{ik_{\mathrm{F}} x}\psi_{r,\alpha}(x).
\label{eq:lrm}
\end{equation}
The Hamiltonian then becomes:
\begin{equation}
H = H_0 + H_{\mathrm{imp}},
\end{equation}
with
\begin{equation}
H_0 = \frac{v_{\mathrm{F}}}{2\pi} \sum_{\alpha} \int_0^L dx 
                             \left(\psi_{l,\alpha}^\dagger(x)
                                   i\partial_x \psi_{l,\alpha}(x) -
                                   \psi_{r,\alpha}^\dagger(x)
                                   i\partial_x \psi_{r,\alpha}(x)\right),
\end{equation}
and
\begin{equation}
H_{\mathrm{imp}} = \lambda \sum_{\alpha, \beta}
           \left(\psi_{l,\alpha}^\dagger(0) + \psi_{r,\alpha}^\dagger(0)\right)
               \vec{\sigma}_{\alpha \beta}
           \Big(\psi_{l,\beta}(0) + \psi_{r,\beta}(0)\Big) \cdot \vec{S}.
\end{equation}

To make progress, it is convenient to pass to a basis of definite parity 
fields ({\em Weyl basis}):
\begin{equation}
\psi_{\mathrm{even},\alpha}(x) = \frac{1}{\sqrt{2}}\left(\psi_{r,\alpha}(x) + 
                                        \psi_{l,\alpha}(-x)\right),
\label{eq:Weyl1}
\end{equation}
an even--parity, right--moving electron field, and
\begin{equation}
\psi_{\mathrm{odd},\alpha}(x) = \frac{1}{\sqrt{2}}\left(\psi_{r,\alpha}(-x) - 
                                        \psi_{l,\alpha}(x)\right),
\label{eq:Weyl2}
\end{equation}
an odd--parity, left--moving field. One should note that the 
assignment of chirality ({\em left/right}) to parity ({\em odd/even}) is 
not intrinsic, but a property of the particular transformations  
(\ref{eq:Weyl1}) and (\ref{eq:Weyl2}). This is analogous to a gauge--fixing 
condition.
In this basis the Hamiltonian takes the form:
\begin{equation}
H = H_0^{\mathrm{odd}} + H_0^{\mathrm{even}} 
                       + H_{\mathrm{imp}}^{\mathrm{even}},
\label{h_parity}
\end{equation}
where
\begin{equation}
H_0^{\mathrm{even}}   = 
-\frac{v_{\mathrm{F}}}{2\pi} \sum_{\alpha} \int_0^L dx 
                                    \psi_{\mathrm{even},\alpha}^\dagger(x)
                                   i\partial_x \psi_{\mathrm{even},\alpha}(x)
\label{Ho.even}
\end{equation}
and
\begin{equation}
H_0^{\mathrm{odd}}  = 
\frac{v_{\mathrm{F}}}{2\pi} \sum_{\alpha} \int_0^L dx 
                                    \psi_{\mathrm{odd},\alpha}^\dagger(x)
                                   i\partial_x \psi_{\mathrm{odd},\alpha}(x)
\label{Ho.odd}
\end{equation}
describe independent relativistic electrons, and
the impurity contribution is now also diagonal:
\begin{equation}
H_{\mathrm{imp}}^{\mathrm{even}} = \lambda \sum_{\alpha, \beta} 
                                        \psi^\dagger_{\mathrm{even},\alpha}(0)
                                        \vec{\sigma}_{\alpha \beta}      
                                         \psi_{\mathrm{even},\beta}(0) \cdot
                                         \vec{S}.
                                         \label{eq:diagonalKondo}
\end{equation}
We recognize $H_K^{\mathrm{even}} \equiv 
              H_0^{\mathrm{even}} + H_{\mathrm{imp}}^{\mathrm{even}}$ 
as the chiral Hamiltonian of the spin-$S$ Kondo model. 

While the even and odd parity channels are decoupled in the Hamiltonian,
they become connected by the twisted boundary conditions (\ref{eq:tbc}):
\begin{equation}
\left(\begin{array}{c} \psi_{{\rm even},\alpha}(L) \\ 
\\
\psi_{{\rm odd},\alpha}(L) \end{array}\right) =
\left(\begin{array}{cc} \cos\phi_\alpha & i\sin\phi_\alpha  \\
\\
-i\sin\phi_\alpha  & \cos\phi_\alpha \end{array}\right)
\left(\begin{array}{c} \psi_{{\rm even},\alpha}(0) \\ 
\\
\psi_{{\rm odd},\alpha}(0) \end{array}\right),
\label{eq.tbc2}
\end{equation}
where in (\ref{eq:lrm}) we have taken $k_F=(2\pi/L)n$, with $n$ an integer.    
However, for the special values $\phi_\alpha=f_\alpha\pi$, where
$f_\alpha$ is an integer,
the matrix in Eq.\ (\ref{eq.tbc2}) reduces to a multiple
of the unit matrix, and the even and odd parity states decouple from each
other entirely.  One can then solve $H_K^{\rm even}$ by the Bethe 
ansatz.\cite{baKondo} Thus, our original problem in
(\ref{eq:Hinitial}) and (\ref{eq:tbc}) has collapsed to an exactly 
solvable problem for $f_\alpha \in Z$, 
consisting of a left-moving odd-parity branch of independent
relativistic electrons, together with a (decoupled) right-moving 
even-parity branch
defined by the 1D Kondo model.
For generic values of $\phi_\alpha$, it is not possible to choose a basis which
renders the Hamiltonian and the boundary conditions simultaneously 
diagonal, suggesting that the model is not 
integrable in 
general. This is in apparent contradiction to
recent claims in the
literature\cite{schlott} about the integrability of the related Anderson ring
threaded by an Aharonov-Bohm flux of arbitrary strength.

 From Eq.\ (\ref{eq:diagonalKondo}), the impurity is seen to couple only to 
the spin current of the electrons, suggesting, via the dynamic spin-charge 
separation in 1D, that the charge persistent 
current is insensitive to the presence of the impurity. Although this  
indeed turns out to be the case---as we shall confirm via a 
Bethe ansatz analysis---some caveats are appropriate at this point: 
First, the persistent current is 
a boundary effect and, as such, could be influenced by 
{\em non-dynamical} selection rules for combining charge and 
spin.\cite{FrojdhJohannesson,cas} Secondly, and possibly reflecting this, a 
magnetic impurity {\em does} affect the charge current of a {\em chiral} 
ring of free electrons (with all electrons moving in the same 
direction).\cite{ZB} In any event, it is instructive to study the exact 
mechanism by which the charge persistent current in the present problem 
avoids any influence from the impurity.  
Moreover, the above is only true - as our exact Bethe ansatz solution shows - 
for $\phi_\uparrow=\phi_\downarrow$.
In general, there is a marked effect of the presence of the magnetic impurity 
on the charge
persistent current.

To carry out this analysis, we first need to consider how to properly  
define a persistent current for relativistic electrons, i.e. for 
electrons with a {\em linear} dispersion.

\section{CHARGE PERSISTENT CURRENT FOR RELATIVISTIC ELECTRONS}
In the usual treatment of independent 1D electrons,\cite{BIL,Cheung} 
the persistent current is obtained
by summing the partial currents $I_n = -(e/\hbar) \partial E_n/\partial \phi$ 
over all occupied levels $n$. This approach clearly fails for relativistic 
electrons since the corresponding 
linear dispersions
\begin{equation}
E_{n_r} = \hbar v_{\mathrm{F}}\frac{2\pi n_r + \phi}{L}, \ \ \ \ 
n_r = 0,1,2,...,n_F
\label{eq:RightLevels}
\end{equation}
and
\begin{equation}
E_{n_l} = \hbar v_{\mathrm{F}}\frac{-2\pi n_l+\phi}{L}, \ \ \ \ n_l = 
1,2,..., n_F
\label{eq:LeftLevels}
\end{equation}
imply that $\partial E_n/\partial \phi = \mathrm{const.}$ for all levels $n$.
(Here, for simplicity, we consider a system of spinless electrons in which the total
number of electrons $2n_F+1$ is odd, with $l\, (r)$ denoting, as
before, a branch of left (right) moving electrons.) 
To recover the known results for the persistent current, we must thus use a
different 
approach.\cite{add} Let us introduce {\em flux-dependent} particle numbers 
\begin{equation}
N_{r/l}(\phi) = \frac{L}{2\pi}[|k_{r/l,F}(\phi)|-|k_{r/l,F}(0)|],
\label{eq:particlenumber}
\end{equation}
where $k_{r/l,F}$ are flux-dependent 
Fermi momenta, associated with the highest occupied level on the 
respective branch.  The Fermi momenta $k_{r,F}$ and $k_{l,F}$ are cutoff dependent, and need
not be equal. 
However,
provided the cutoffs are chosen independent of $\phi$, $N_{r/l}(\phi)$
are insensitive to the cutoffs, and describe the physical 
response of the system to an ABC flux.  The persistent current 
is then
\begin{equation}
I(\phi) = -\frac{e v_F}{L}[N_r(\phi)-N_l(\phi)].
\label{eq:current}
\end{equation}
It should be pointed out that the charge velocity $v_F$ is in general
subject to renormalization due to electron-electron interactions. 

With the choice of representative levels in (\ref{eq:RightLevels}) and 
(\ref{eq:LeftLevels}) (note in particular that the zero mode 
is assigned to {\em one} branch only) 
it is easy to 
verify that (\ref{eq:particlenumber}) and (\ref{eq:current}) exactly reproduce
the known 
result for
the persistent current of an odd number of
spinless 1D electrons.\cite{Cheung} Our construction, introduced here 
{\em ad hoc}, can trivially be extended to spinful particles and 
put on a firm basis by a proper analysis of the cutoff procedure for
1D relativistic electrons in the presence of ABC fluxes.\cite{EJS}  
In short, a flux-dependent particle number as in (\ref{eq:particlenumber})
is the trade-off that guarantees that physical observables remain independent
of the choice of cutoff which bounds the spectrum of a finite system from below.

Given (\ref{eq:current}), the problem is 
now reduced to calculating how the effective particle numbers depend 
on the flux and the coupling of the electrons to the magnetic impurity.
For this, we turn to a finite-size Bethe ansatz analysis.  

\section{FINITE--SIZE BETHE ANSATZ}
To obtain the flux dependent particle numbers for a finite ring, 
we apply
the techniques of the {\em Bethe ansatz} for finite systems,
developed previously for the 1D Hubbard
model.\cite{we}
As pointed out above, our model is only integrable for 
$\phi_\alpha=f_\alpha\pi$, with 
$f_\alpha$ an integer.  For $f_\alpha\in Z$, the 
nested Bethe ansatz equations which diagonalize $H$ in (\ref{h_parity})
are 
\begin{equation}
Lk_{n_{l}}= -2\pi n_{l} + f_c\pi 
+\left(\frac{2M_{\rm odd}}{N_{\rm odd}} - 1\right)f_s\pi
+ \frac{2\pi}{N_{\rm odd}}
\sum_{\delta=1}^{M_{\rm odd}} J_\delta,
\label{eq:leftmovers}
\end{equation}
\begin{equation}
Lk_{n_r} = 2 \pi n_{r} + f_\downarrow\pi + \sum_{\gamma = 1}^{M_{\rm even}} 
                 \left[\Theta(2\Lambda_\gamma - 2) - \pi \right],
\label{eq:holonBAE}
\end{equation}
\begin{equation}
N_{\rm even}
 \Theta(2\Lambda_\gamma - 2) + \Theta(2\Lambda_\gamma) = 2 \pi I_\gamma
+(f_\uparrow - f_\downarrow)\pi
 + \sum_{\delta = 1}^{M_{\rm even}} \Theta(\Lambda_\gamma -\Lambda_\delta ),
\label{eq:spinonBAE}
\end{equation}
where $k_{n_l}$ are the pseudomomenta characterizing
the  $N_{\rm odd}$ odd-parity {\em left} movers 
which decouple from the impurity, 
$M_{\rm odd}$ of which have spin down,
and $k_{n_r}$ are pseudomomenta characterizing
the $N_{\rm even}$ even-parity {\em right} movers, 
$M_{\rm even}$ of which have spin down. The numbers $n_l, n_r,
I_{\gamma}$ and $J_{\delta}$ take integer or half-odd integer values
depending on the values of $M_{even/odd}$ and $N_{even/odd}$ (see
below),
while $\{\Lambda_\gamma$, $\gamma=1,\cdots,M_{\rm even}\}$
are a set of auxiliary variables known as spin-rapidities. 
The scattering phase shifts are given by $\Theta(x) = 2\tan^{-1}(x/c)$,
with $c=2\lambda/(1-3\lambda^2/4)$.
We have also defined $f_{c,s} = (f_\uparrow \pm f_\downarrow)/2$. 
Eq.\ (\ref{eq:leftmovers}) simply gives the quantum numbers of free, chiral
electrons, written in the Bethe ansatz basis.
The Bethe ansatz equation (\ref{eq:holonBAE}) describes the 
charge degrees of freedom in the even channel (holons), while Eq.\ 
(\ref{eq:spinonBAE}) describes
the spin degrees of freedom in the even channel (spinons). 
Eqs.\ (\ref{eq:holonBAE}) and (\ref{eq:spinonBAE}) differ from the Bethe
ansatz equations
derived previously for the Kondo model\cite{baKondo} only by the addition
of the ABC fluxes $\phi_\alpha=f_\alpha\pi$.

Let us consider for the moment the case of a
spin-independent flux $\phi_\uparrow = \phi_\downarrow = \phi$, corresponding
to the case where only a magnetic flux threads
the ring and there is no charged string passing through the ring (AB flux only).
The persistent current is an odd function of $\phi$ by symmetry,\cite{by}
and is analytic, except at values of $\phi$ corresponding to level
crossings.
We are interested in the persistent current for small values of the 
AB flux.  Choosing the total numbers of both up- and down-spin electrons
to be odd excludes a level crossing at $\phi=0$.
The leading mesoscopic behavior of the persistent current is then
\begin{equation}
I(\phi) = -D_c \phi/L + {\cal O}(\phi^3/L^3),
\label{stiff}
\end{equation}
where $D_c$ is the charge stiffness.  Eq.\ (\ref{stiff}) holds on general 
grounds independent of whether the model is integrable or not.

The choice of quantum numbers $\{n_l, J_\delta, n_r, I_\gamma\}$ 
specifies the quantum state of the system.  Generically, there are one or
more level crossings\cite{casajm} between $f=0$ and $f=1$.  
To determine the charge stiffness, however,
we only need to consider the state which evolves 
adiabatically from the ground state at $f=0$ as $\phi$ is increased.  This
state is given by $M_{\rm even/odd} = (N_{\rm even/odd}+/- 1)/2$, 
(with $N_{\rm even/odd}$ odd for simplicity), with
integer-spaced quantum numbers $\{n_l, J_\delta,
n_r, I_\gamma\}$ in the symmetric
ranges $-(N_{\rm odd}-1)/2 \leq n_l \leq (N_{\rm odd}-1)/2$,
$-(M_{\rm odd} - 1)/2 \leq J_\delta \leq (M_{\rm odd} - 1)/2$,
$-(N_{\rm even}-1)/2 \leq n_r \leq (N_{\rm even}-1)/2$, and 
$-(M_{\rm even} - 1)/2 \leq I_\gamma \leq (M_{\rm even} - 1)/2$.
The quantum numbers of the even-parity sector are 
the same as those of the Kondo model with periodic
boundary conditions.\cite{baKondo}

Given a set of 
spin rapidities $\Lambda_\gamma$ satisfying Eq.\ (\ref{eq:spinonBAE}), 
we may calculate the sum in Eq.\ (\ref{eq:holonBAE}), and thus
the momenta $k_{n_r}$ are determined.  One sees immediately that the
total scattering phase shift of the dressed magnetic impurity is {\em
independent} of $f$, so that $N_r=-N_l=f/2$. 
In addition, the charge velocity $v_F$
is unrenormalized by
interactions in this model.\cite{baKondo}
The charge stiffness may be evaluated from Eqs.\ (\ref{eq:current}) 
and (\ref{stiff}) as a finite difference
$D_c = -LI(f=1)/\pi=ev_F/\pi+ {\cal O}(L^{-2})$. 
This gives a lower bound to the charge stiffness, since an avoided level
crossing in the nonintegrable regime $0<\phi<\pi$ cannot be excluded.
However, it is difficult to imagine on physical grounds how the magnetic
impurity could enhance the persistent current, so we expect that this lower
bound is an equality. 
The persistent current for small $\Phi$ is thus
\begin{equation}
I= -\frac{ev_F}{L}\frac{2\Phi}{\Phi_0},
\label{eq:final}
\end{equation}
which is identical to the result for free electrons. Eq.\ (\ref{eq:final})
indicates that
spin-charge separation holds even at the mesoscopic scale in this model.

The analysis of the Bethe ansatz equations 
(\ref{eq:leftmovers}-\ref{eq:spinonBAE}) in the case of general spin--dependent
fluxes (however, still satisfying $\phi_\alpha=f_\alpha\pi$ with $f_\alpha$ an
integer) is more involved and will be presented elsewhere.\cite{EJS} 
However, an analysis of the equations
(\ref{eq:leftmovers}-\ref{eq:spinonBAE}) 
in the limiting cases
$c\rightarrow0$ and $c\rightarrow\infty$ indicates that the charge
persistent current is markedly affected by the magnetic impurity when 
spin-dependent fluxes are present.  This is because the AC effect induces
a charge persistent current if the numbers of up and down spin electrons are
not equal.  As $c$ is increased from 0 to $\infty$, the impurity screens
exactly one electron spin, so the effective numbers of mobile up and down spin 
electrons are not equal in general.

\section*{ACKNOWLEDGEMENTS}

This work was supported in part by a bilateral grant from the Finnish
Academy and the Deutscher Akademischer Austauschdienst.
HPE is an Alexander--von--Humboldt Fellow of the Finnish Academy. He thanks
the Department of Physics of the University of Arizona 
and the Institute of Theoretical Physics at Chalmers and G\"oteborg 
University for kind hospitality during crucial stages of this 
work. HJ acknowledges financial support from the Swedish Natural Science 
Research Council.


\begin{thebibliography}{9}

\bibitem{review} For a review,
see e.g. Science {\bf 281}, 540 (1998), and references therein.
\bibitem{li} J.\ Li, W.-D.\ Schneider, R.\ Berndt, and R.\ Delley,
Phys.\ Rev.\ Lett.\ {\bf 80}, 2893 (1998).
\bibitem{yacoby} A.\ Yacoby, M.\ Heilblum, D.\ Mahalu, and H.\
Shtrikman, Phys.\ Rev.\ Lett.\ {\bf 74}, 4047 (1995).
\bibitem{by}N.\ Byers and C.\ N.\ Yang, Phys.\ Rev.\ Lett.\ {\bf 7}, 46 (1961).
\bibitem{BIL} M. B\"{u}ttiker, Y. Imry, and R. Landauer, 
Phys. Lett. {\bf A96}, 365 (1983). 
\bibitem{Cheung}
H.-F. Cheung {\em et al.}, Phys.\ Rev.\ B {\bf 37}, 6050 (1988).
\bibitem{mbcs} M.\ B\"uttiker and C.\ A.\ Stafford, Phys.\ Rev.\ Lett.\
{\bf 76}, 495 (1996).
\bibitem{baKondo}For a recent review, see: N.\ Andrei, Integrable Models in 
Condensed Matter Physics, in {\em Series on Modern Condensed Matter Physics -
Vol. 6}, 458-551 (World Scientific, Singapore, 1992), Eds. S.\ Lundquist,
G.\ Morandi, and Yu Lu.
\bibitem{FrojdhJohannesson} P. Fr\"ojdh and H. Johannesson, Phys.\ Rev.\ Lett.
{\bf 75}, 300 (1995); 
Phys.\ Rev.\ B {\bf 53}, 3211 (1996). 
\bibitem{cas} C. A. Stafford, Phys. Rev. B {\bf 48}, 8430 (1993).
\bibitem{schlott} P.\ Schlottmann and A.\ A.\ Zvyagin, Phys.\ Lett.\ A
{\bf 231}, 109 (1997).
\bibitem{ZB} A. A. Zvyagin and T. V. Bandos, Low Temp. Phys. {\bf 20}, 
222 (1994).  
\bibitem{add} For alternative methods 
for determining the persistent
current for a system with linear spectrum (exploiting bosonization),
see e.g. D. Loss, Phys. Rev. Lett. {\bf 69}, 343
(1992), or A. O. Gogolin and N. V. Prokof'ev, Phys. Rev. B {\bf 50},
4921 (1994). 
\bibitem{EJS} H.-P. Eckle, H. Johannesson and C. A. Stafford, 
(unpublished).
\bibitem{we}F.\ Woynarovich and H.-P.\ Eckle, J.\ Phys.\ A: Math.\ Gen.\ 
{\bf 20}, L443 (1987); F.\ Woynarovich, J.\ Phys.\ A: Math.\ Gen.\ {\bf 22},
2615 (1989).
\bibitem{casajm} N.\ Yu and M.\ Fowler, Phys.\ Rev.\ B {\bf 45}, 11795 (1992);
C.\ A.\ Stafford and A.\ J.\ Millis, Phys.\ Rev.\ B {\bf 48}, 
1409 (1993). 
\end{thebibliography}
\end{document}